\documentclass[twocolumn,showpacs,preprintnumbers]{revtex4}

\input{epsf}

\def\JHEP{{\it JHEP} }

\def\PL{{\it Phys. Lett.} }
\def\PR{{\it Phys. Rev.} }

\def\frac#1#2{{\textstyle{{#1}\over {#2}}}}

\def\lsim{\mathrel{\rlap{\lower4pt\hbox{\hskip1pt$\sim$}}
    \raise1pt\hbox{$<$}}}
\def\gsim{\mathrel{\rlap{\lower4pt\hbox{\hskip1pt$\sim$}}
    \raise1pt\hbox{$>$}}}
\def\sqr#1#2{{\vcenter{\vbox{\hrule height.#2pt
         \hbox{\vrule width.#2pt height#1pt \kern#1pt
         \vrule width.#2pt}
         \hrule height.#2pt}}}}

\def\beq{\begin{equation}}
\def\eeq{\end{equation}}
\def\beqa{\begin{eqnarray}}
\def\eeqa{\end{eqnarray}}

\begin{document}
\bibliographystyle {unsrt}
\newcommand{\pa}{\partial}
\title{ Tachyonic Inflation in the Braneworld Scenario}

\author{M. C. Bento$^{1,2}$, O. Bertolami$^1$ and A.A. Sen$^3$}

\vskip 0.2cm

\affiliation{$^1$  Departamento de F\'\i sica, Instituto Superior T\'ecnico \\
Av. Rovisco Pais 1, 1049-001 Lisboa, Portugal}

\vskip 0.2cm

\affiliation{$^2$  Centro de F\'{\i}sica das 
Interac\c c\~oes Fundamentais, Instituto Superior T\'ecnico}

\vskip 0.2cm

\affiliation{$^3$  Centro Multidisciplinar de Astrof\'{\i}sica, 
Instituto Superior T\'ecnico}

\vskip 0.2cm

\affiliation{E-mail addresses: bento@sirius.ist.utl.pt; 
orfeu@cosmos.ist.utl.pt; anjan@x9.ist.utl.pt}

\vskip 0.5cm

\date{\today}

\begin{abstract}

We consider cosmological inflation driven by the rolling tachyon 
in the context of the braneworld scenario. We show that sufficient inflation 
consistent with the observational constraints can be achieved for well defined 
upper limits on the five-dimensional mass scale, string mass scale and 
the string coupling for the bosonic string.     
 
\vskip 0.5cm
 
\end{abstract}

\pacs{98.80.Cq, 98.65.Es \hspace{2cm}Preprint DF/IST-9.2002}

\maketitle

\section{Introduction}

The inflationary scenario offers the attractive possibility of resolving many
puzzles of the standard Hot Big Bang cosmology. The crucial feature of most 
inflationary models is the period of ``slow-roll'' evolution of a scalar
field (the ``inflaton'') during which the potential energy dominates the
kinetic energy  and the universe undergoes a period of exponential expansion.
Although particle physics, in particular string theory, provides several very
weakly coupled scalar fields which are natural inflaton candidates, at
present there exists no clearly preferred  inflationary model; it is
therefore interesting to explore
new possibilities for the inflationary scenario in the early universe.

The study of non-BPS objects such as non-BPS branes, brane-antibrane
configurations and spacelike branes has recently attracted great attention
given its implications for string/M-theory and cosmology. The conjecture that
the classical decay  of unstable D-branes leads, in string theory, to a
pressureless gas with non-vanishing energy density, and that the vacuum of
the tachyon effective action describes a 
configuration with no D-branes, so that around this minimum there are no
physical open string excitations \cite{Sen}, has quickly lead to the idea 
of a tachyon cosmology \cite{Gibbons}. In this scenario,  the main idea is
to consider the coupling with gravity by adding the Einstein-Hilbert action
to the tachyon effective action \cite{Gibbons}. 
From this perspective, it is quite natural to consider the possibility of
inflation driven  by the rolling tachyon.

Recently, several authors have investigated the cosmological implications
of a tachyon rolling  down its potential, which depends on the underlying
bosonic or supesymmetric theory \cite{FT,SW,FKS,KL,sami,ddds}, to its 
ground state.
Tachyonic inflation \cite{fien, paddy} has also been studied using
phenomenological potentials that have not been derived from  string theory
and can be related to the so called ``k-Inflation'' \cite{mukh}.

It has been shown  by Kofman and Linde \cite{KL} that it is very difficult to
find realistic inflationary models  in the context of
string theory tachyon condensation, as it leads to unacceptably large metric
perturbations. Also,  inflation in these theories occurs only at
super-Planckian values of the brane energy density, making the effective 
four-dimensional gravity theory unreliable near the top of 
the potential \cite{KL}.
 
A natural extention of this proposal is to consider the tachyon
as a degree of freedom on the visible three-dimensional brane. 
Hence, the Einstein-Hilbert action would
describe  dynamics on the bulk, which is most often considered to be a
five-dimensional AdS spacetime background \cite{Mukohyama}. This proposal
implies that, on the brane, the dynamics would be described by the modified
Einstein equations \cite{Shiromizu}. In this work, we will show that these
underlying assumptions allow for an inflationary period in which all known
observational constraints are successfully met. As we shall see, an important 
advantage  of our proposal is that this period of inflation occurs at the top
of the tachyonic potential emerging from the tachyonic effective theory. Also,
the brane energy density remains sub-Planckian allowing us to work with the
effective four-dimensional gravity theory. Moreover, we find 
constraints on the fundamental string mass scale $M_{s}$ and the 
five-dimensional Planck mass, $M_{5}$.

We will show how our approach allows avoiding, to some extent, difficulties
with tachyonic inflation \cite{KL}, even though there is still the serious
problem of reheating whose solution does require a better understanding of
the couplings of the tachyon to matter \cite{Cline}.   

\section{Tachyonic Inflation}

The effective field theory action for the tachyon field $T$ on a $D3$ brane 
computed  in the bosonic String Field Theory (SFT), around the top of 
the potential, is given by \cite{BSFT} 

\beq
S_B = \tau_3 \int d^4 x\exp(-T)[l_s^2 \partial_{\mu} T \partial^{\mu} T 
+ (1 + T)]~~,
\label{eq:sft}
\eeq
where the normalization factor $\tau_3$ is the D3 brane tension 

\beq
\tau_3 = {M_{s}^{4}\over{(2\pi)^{3}g_{s}}}~~,
\label{eq:tau}
\eeq
$g_s$ is the string coupling, and the string mass scale is given by 
$M_s^2 = 1/l_s^2$.

The action given by Eq. (\ref{eq:sft}) is exact up to second order in $\partial_{\mu} T
\partial^{\mu} T$
and accounts for the effects of all open string modes. Notice that, with our 
conventions, the tachyon field $T$ is dimensionless and hence the potential
$V(T)=\tau_3 (1+T)\exp(-T)$ has mass dimension four. Moreover, due to the
factor $\exp(-T)$, the kinetic term has a nonstandard form; it is, however,
possible  to eliminate this factor via a field redefinition 

\beq
\phi = \exp(-T/2)~~,
\label{eq:red}
\eeq
and hence obtain a canonical kinetic term; in this case, the potential becomes 

\beq
V(\phi) = -\tau_{3}\phi^{2}\log(\phi^2/e)~~.
\label{eq:newv}
\eeq

Now the ``stable vacuum'' to which the tachyon condenses is at 
$T\rightarrow\infty$ (or at $\phi=0$) and, for this vacuum,
$V^{''}\rightarrow\infty$ and  the tachyon  acquires an infinite mass.
Hence, with a standard kinetic term for $\phi$, we have a field theory
for the tachyon with the property that it rolls down to its stable
vacuum and condenses, disappearing from the spectrum as it acquires an 
infinite mass. This is the manifestation of the disappearance of the whole
tower of the open string fields  and is consistent with Sen's
conjecture.

One can also write a closed form expression for the action Eq. (\ref{eq:sft}), 
incorporating all the higher powers of $\partial_{\mu}T$. The effective 
tachyon field action in Born-Infield form, ignoring the second and 
higher derivative terms, is described in the presence of gravity by 

\beq
S_{BI} = \tau_3 \int d^4 x \sqrt{-g}~V(T) 
\sqrt{1 + l_{s}^{2} f(T) \partial_{\mu} T \partial^{\mu} T}~~.
\label{eq:BI}
\eeq

For small $\partial_{\mu}T$, one can treat the action Eq. (\ref{eq:sft}) as an
expansion of action Eq. (\ref{eq:BI}) with $V(T)=(1+T)\exp(-T)$ and
$f(T)=2(1+T)^{-1}$. It should, however, be noted that late time evolution of
the tachyon field, in either case, does not seem to match the asymptotic
behaviour conjectured by Sen. However, we are interested in the early time
evolution of the tachyon field, when it is slightly displaced from the top of
the potential at $T=0$ (or $\phi=1$), where the time derivatives of the
tachyon turn out to be small, and one can safely assume either action
 (\ref{eq:sft}) or (\ref{eq:BI}). We shall proceed with the action given by Eq. 
(\ref{eq:sft}) 
and the field redefinition given by Eq. (\ref{eq:red}) but it should be noted 
that, if one uses the action of Eq. (\ref{eq:BI}), with $V(T)$ and
 $f(T)$ given 
above, within the slow-roll approximation, the results are the same.

We now show that the potential (\ref{eq:newv}), within the five-dimensional 
brane scenario, leads to a successful inflationary model.

In the five-dimensional ADS braneworld scenario, the Friedmann equation on the 
visible brane becomes \cite{Shiromizu,Binetruy}:

\beq
H^2 = {\Lambda \over 3} +  \left({8 \pi \over 3 M_P^2}\right) \rho
+ \left({4 \pi \over 3 M_5^3}\right)^2 \rho^2 + {\epsilon \over a^4},
\label{eq:H2}
\eeq
where 
\beq
M_P = \sqrt{{3 \over 4 \pi}} {M_5^3 \over \sqrt{\lambda}}~~,
\label{eq:MP}
\eeq
$\lambda$ is the brane tension, which relates 
the four and five-dimensional Planck scales 
and the four and five-dimensional cosmological constants via
the relationship

\beq
\Lambda = {4 \pi \over M_5^3} \left(\Lambda_5 + {4 \pi \over 3 M_5^3}~
\lambda^2 \right)~~.
\label{eq:Lam}
\eeq

Assuming that, as required by observations, the cosmological constant 
is negligible in the early universe and since the last term 
in Eq. (\ref{eq:H2}) rapidly becomes unimportant after inflation sets in, 
the Friedmann equation becomes

\beqa
H^2& =& {8 \pi \over 3 M_P^2} \rho \left[1 + {\rho \over 2 \lambda}\right]\\
&=&\left( 4\pi\over 3 M_5^3\right)^2 \rho^2\left(1+{\rho_c\over \rho}\right)~,
\label{eq:H22}
\eeqa
with

\beq
\rho_c={3\over 2\pi}{ M_5^6\over M_P^2}~~.
\eeq

Hence, the new term in $\rho^2$ is dominant at high energies, compared
to $\lambda^{1/4}$, i.e. $\rho>\rho_c$, but quickly decays at lower energies.

Finally, we shall assume that the scalar field is confined to the brane, so
that its field equation has the standard form

\beq
\ddot \phi + 3 H \dot \phi + {d V\over d \phi} = 0~.
\eeq

The number of e-folds during inflation is given by $N =
\int_{t_{\rm i}}^{t_{\rm f}} H dt$, which becomes \cite{Maartens}

\beq
\label{eq:N}
N \simeq - (8\tau_{3}l_{s}^{2}){8\pi  \over M_{P}^2}
\int_{\phi_{\rm i}}^{\phi_{\rm f}}{V\over V'} 
\left[ 1+{V \over 2\lambda}\right]  d\phi~~,
\eeq
in the slow-roll approximation. We see that, as a result of the modification
in the Friedmann equation,
the expansion rate is increased, at high energies,  by a factor $V/2\lambda$.

One can also define the two slow-roll parameters
\cite{Maartens}

\begin{eqnarray}
\label{eq:epsilon}
\epsilon &\equiv&{M_{P}^2 \over 16\pi}{1\over{8\tau_{3}l_{s}^{2}}} 
\left( {V' \over V}\right)^2  {1+{V/ \lambda}\over(2+{V/\lambda})^2}~~,\\
\label{eq:eta}
\eta &\equiv& {M_{P}^2\over{8\pi}}{1\over{8\tau_{3}l_{s}^{2}}} 
{V'' \over V}   {1 \over 1+{V/ 2 \lambda}}~~.
\end{eqnarray}

Notice that both parameters are
suppressed by an extra factor $\lambda/V$ at high energies and that, at low
energies, $V\ll\lambda$, they reduce to the standard
form. The extra factor $1/ 8\tau_{3}l_{s}^{2}$ appears in 
the expressions for $N,\epsilon$ and $\eta$ due to the presence of the 
factor $\tau_{3}l_{s}^{2}$ in the kinetic energy term in Eq. (\ref{eq:sft}).

The value of $\phi$ at the end of inflation can be obtained 
from the condition

\beq
{\rm max}\{\epsilon(\phi_F),|\eta(\phi_F)|\}= 1~~.
\label{eq:phif}
\eeq

We shall use the high energy approximation ($V \gg \lambda$) in our 
calculations which, for the potential Eq. (\ref{eq:newv}), implies

\beq
M_{5} < \left({1\over{6\pi^2 g_{s}}}\right)^{1/2}\left({M_{5}\over{M_{s}}}\right)^{-2}M_{P}~~.
\label{eq:M5b}
\eeq

For our model, the slow-roll parameters are given by

\beq
\epsilon = -{2a(1+\log(\phi^{2}/e))^{2}\over{8\pi\phi^{4}
\log^3(\phi^{2}/e)}}~~,
\label{eq:eps}
\eeq

\beq
\eta = -{2a (6+2\log(\phi^{2}/e))\over{8\pi\phi^{4}\log^2(\phi^{2}/e)}}~~,
\label{eq:etaf}
\eeq
where 

\beq
a \equiv 6\pi^{5} g_{s}^{2}(M_{5}/M_{s})^{6}~~.
\label{eq:a}
\eeq
\begin{figure}[t]
\centering
\leavevmode \epsfysize=5cm \epsfbox{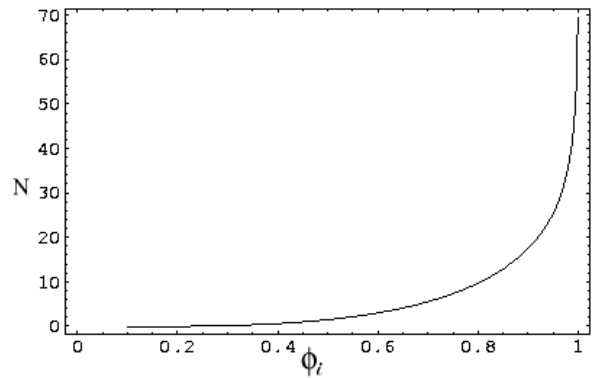}\\
\vskip 0.1cm
\caption{Inflation occurs  primarily for initial values of the 
inflaton field near $\phi= 1$,  corresponding  to  the top of the potential.}
\label{fig:figure1}
\end{figure}

\begin{table*}[t]
\caption[par]{\label{tab:par} Relevant physical quantities 
 for the tachyonic inflationary potential of Eq.~(\ref{eq:sft}), for different values of  $a$.}
\begin{ruledtabular}
\begin{tabular}{|c|c|c|c|c|c|c|c|}
\hline
$a$  & $\phi_{f}$  & $N$  & $\phi_{\star}$  & $n_{s}$ & $r_{s}$ & $M_{5}$ 
(upper bound) & $M_{s}$ (upper bound)\\
\hline
0.28 & 0.24 & 70 & 0.996 & 0.83 & 9.7$\times 10^{-5}$ & 5.6$\times 10^{-3}
M_{P}$ & 1.3$\times 10^{-7}M_{P} $\\
\hline
0.23 & 0.22 & 85 & 0.990 & 0.86 & 4.5$\times 10^{-4}$ & 7.3$\times 10^{-3}
M_{P}$ & 2.6$\times 10^{-7}M_{P}$\\
\hline
0.21 & 0.22 & 90 & 0.988 & 0.86 & 6.7$\times 10^{-4}$ & 7.8$\times 
10^{-3}M_{P}$ & 3.0$\times 10^{-7}M_{P}$\\
\hline
0.20 & 0.22 & 96 & 0.986 & 0.87 & 9.7$\times 10^{-4}$ & 8.3$\times 
10^{-3}M_{P}$ & 3.5$\times 10^{-7}M_{P}$\\
\hline
0.19 & 0.21 & 103 & 0.982 & 0.88 & 1.4$\times 10^{-3}$ & 8.8$\times 
10^{-3}M_{P}$ & 4.1$\times 10^{-7}M_{P}$\\
\hline
0.17 & 0.21 & 110 & 0.978 & 0.89 & 2.0$\times 10^{-3}$ & 9.3$\times 
10^{-3}M_{P}$ & 4.5$\times 10^{-7}M_{P}$\\
\hline
0.15 & 0.20 & 129 & 0.966 & 0.90 & 4.1$\times 10^{-3}$ & 1.1$\times 
10^{-2}M_{P}$ & 6.2$\times 10^{-7}M_{P}$\\
\hline
0.13 & 0.19 & 155 & 0.947 & 0.92 & 8.1$\times 10^{-3}$ & 1.2$\times 
10^{-2}M_{P}$ & 8.0$\times 10^{-7}M_{P}$\\
\end{tabular}
\end{ruledtabular}
\end{table*}


The expression for number of e-foldings now becomes

\beq
N \simeq {8\pi\over{4a}}\int_{\phi_{i}}^{\phi_{f}}{\phi^{3}
\log^2(\phi^{2}/e)\over{1+\log(\phi^{2}/e)}}d\phi~~.
\label{eq:newN}
\eeq
For $N\simeq 70$, the minimum number of e-foldings required to solve 
different cosmological problems, Eqs.~(\ref{eq:newN}) and (\ref{eq:phif}) 
lead to $a\simeq 0.279$ and $\phi_f\simeq 0.236$, assuming $\phi_{i}=0.999$. Indeed, we 
have checked that sufficient inflation can easily be achieved if the field 
starts very near the top of the potential, at $\phi\simeq 1$ (See Figure 1).

We now turn to the constraints imposed by  the observed CMB anistropies.
The amplitude of scalar perturbations is given by \cite{Maartens}

\beq
\label{eq:As}
A_{s}^2 \simeq  8\tau_{3}l_{s}^{2}{512\pi\over 75 M_P^6}
 {V_\star^3\over V_\star^{\prime2}}
\left[ 1 + {V_\star \over 2\lambda} \right]^3~~,
\eeq
with $V_\star=V(\phi_\star)$, where  $\phi_\star$ is the value of $\phi$ when
scales corresponding to
large-angle CMB anisotropies, as observed by COBE, left the
Hubble radius during inflation i.e. approximately after 
55 e-foldings. For  the potential of Eq. (\ref{eq:newv}), again 
using the high energy
approximation, the above equation becomes 

\beq
A_{s}^{2} \simeq {512\pi^{3}\over{4800}}{g_{s}\phi_{\star}^{10}\over{a^{3}}}
{\log^6(\phi_{\star}^{2}/e)\over{\log^2(\phi_{\star}^{2}/e +1)}}~~.
\label{eq:Asn}
\eeq

For our model, with $N_\star\approx 55$,
$\phi_i=0.999$ and $a=0.279$, we get, using Eq.~(\ref{eq:newN})

\beq
\phi_\star \simeq 0.996~~.
\label{eq:phistar2}
\eeq
Inserting the above values of $a$ and $\phi_{\star}$ into Eq. (\ref{eq:Asn}) 
and using the fact that the  observed value
from COBE is $A_{s}\simeq 2\times 10^{-5}$, we get a constraint on $g_{s}$,   
namely $g_{s}\simeq 1.5\times 10^{-16}$. Using these values of $g_{s}$ and $a$ 
in Eqs. (\ref{eq:M5b}) and (\ref{eq:a}), we  get the following bounds on 
$M_{5}$ and $M_{s}$:

\begin{eqnarray}
M_{5} \lsim 5.6 \times 10^{-3} M_{P}~~,~~
M_{s} \lsim  1.3 \times 10^{-7} M_{P}~~.
\end{eqnarray} 

The scale-dependence of the perturbations is described by the
spectral tilt \cite{Maartens}

\beq
n_{s} - 1 \equiv {d\ln A_{s}^2 \over d\ln k} \simeq -6\epsilon + 2\eta~~,
\label{eq:ns}
\eeq
where the slow-roll parameters are given in Eqs.~(\ref{eq:epsilon})
and (\ref{eq:eta}). 

The ratio between the amplitude of tensor and scalar
perturbations is given by \cite{Langlois}

\beq
r_s=4\pi \left({A_{t} \over A_{s}}\right)^2 \simeq {3 M_P^2\over 4}
\left({V^\prime\over V} \right)^2 {2 \lambda\over V}~~.
\label{eq:r}
\eeq 
In our model, under the high energy approximation, for $N=70$, we get

\beq
n_{s} \simeq 0.825~,
\label{eq:nsm}
\eeq
and 
\beq
r_s \simeq 9.7 \times 10^{-5}~,
\label{eq:rm}
\eeq
which are
within the bounds resulting from the 
latest CMB data from BOOMERANG \cite{Boom},  MAXIMA \cite{Maxima} and 
DASI \cite{DASI}, namely

\beq
0.8 < n_s < 1.05~~~, \qquad
r_s < 0.3~~.
\label{eq:observations}
\eeq

Repeating  the above calculations  for other values of $a (N)$, 
we obtain Table 1.
We see that, as $a$ decreases, $\phi_f$ decreases and $N$, $n_s$, $r_s$ and 
$M_5$ increase. In all cases, the string remains very weakly coupled 
($g_{s}\sim 10^{-16}$) and the string energy density $\tau_{3}$  remains 
sufficiently below the Planck scale ($\tau_{3}<10^{-12}M_{P}^4$) so that one 
can safely use the low energy four-dimensional gravity theory.

\section{Conclusions}

In this work, 
we have shown that successful tachyonic inflation can be achieved 
in the context of a five-dimensional ADS braneworld scenario. We have shown 
the bosonic SFT action around the top of the tachyon potential does allow a 
sufficiently long period of inflation provided the string remains very weakly 
coupled ($g_{s}\sim 10^{-16}$) and the string energy density $\tau_{3}$  
is sufficiently below the Planck scale ($\tau_{3}<10^{-12}M_{P}^4$) 
to render the low-energy four-dimensional gravity theory description reliable. 
Our setup leads to upper bounds on the string scale, typically 
$M_s \lsim 8 \times 10^{-7}~M_P$, and on the five-dimensional scale, 
$M_5 \lsim 10^{-2}~M_P$ (values of this order seem to be a common feature of 
inflation on the brane, as they arise in chaotic inflation \cite{Bento1} as 
well as in supergravity models \cite{Bento2}).  

We emphasize that, in our formulation, 
inflation takes place very close the top of 
the tachyon potential, thus avoiding 
the recently identified problem of formation 
of caustics with multi-valued regions for scalar Born-Infeld theories with 
arbitrary runaway potentials, meaning that high order spatial derivatives 
of $T$ become divergent \cite{Felder}.

\vskip 0.5cm

\begin{acknowledgments}

\noindent
M.C.B. and  O.B. acknowledge the partial support of Funda\c c\~ao para a 
Ci\^encia e a Tecnologia (Portugal)
under the grant POCTI/1999/FIS/36285. The work of A.A. Sen is fully 
financed by the same grant. 

\end{acknowledgments}



\end{document}